\documentclass[12pt,twocolumn,a4paper]{article}

\usepackage{times}
\usepackage{graphicx}

\title{Relativistic space-time positioning: principles and strategies.}
\author{Angelo Tartaglia\thanks{E-mail: angelo.tartaglia@polito.it}\\ \\ \small{Department of Applied Science and Technology, Politecnico di Torino, and INFN, Italy} }

\begin{document}
\label{paper:tartaglia}

\maketitle

\begin{abstract}
Starting from the description of space-time as a curved four-dimensional manifold,  null Gaussian coordinates systems as appropriate for relativistic positioning will be discussed. Different approaches and strategies will be reviewed, implementing the null coordinates with both continuous and pulsating electromagnetic signals. In particular, methods based on purely local measurements of proper time intervals between pulses will be expounded and the various possible sources of uncertainty will be analyzed. As sources of pulses both artificial and natural emitters will be considered. The latter will concentrate on either radio- or X ray-emitting pulsars, discussing advantages and drawbacks. As for artificial emitters, various solutions will be presented, from satellites orbiting the Earth to broadcasting devices carried both by spacecrafts and celestial bodies of the solar system. In general the accuracy of the positioning is expected to be limited, besides the instabilities and drift of the sources, by the precision of the local clock, but in any case in long journeys systematic cumulated errors will tend to become dominant. The problem can be kept under control properly using a high level of redundancy in the procedure for the calculation of the coordinates of the receiver and by mixing a number of different and complementary strategies. Finally various possibilities for doing fundamental physics experiments by means of space-time topography techniques will shortly be presented and discussed.
\end{abstract}

\section{Introduction}

The problem of positioning is as old as the history of wandering of mankind especially by see. Since the oldest times the problem was tackled looking at the sky and associating the observation with time measuring. Initially time was determined using the rotation of the earth as a clock and one had to wait until the 18th century for the invention of the chronometer to reach an accuracy appropriated to the development of modern technological societies.

In our days the global positioning on earth and close to it is obtained by means of global positioning systems. The first and most used system, starting to be deployed at the end of the '70's of the last century, is named after its acronym GPS. A similar one developed by the former Soviet Union is GLONASS; Europe has started the deployment of its own Galileo system, which at the moment has three satellites in the sky. China is planning to build its global navigation and positioning system, named Bei Dou (North star). India and Japan are also planning to develop national systems, and others are also considering the possibility to do the same. The reason for this vast interest is mainly political, since both GPS and GLONASS are under military control, even when they are used for civilian purposes. In any case all these systems, deployed or under implementation, are such that a continuous control and intervention from the ground is needed.

GPS, which is a sort of an archetype of all current positioning systems, is like a chimera made of different pieces. Basically, even using a constellation of 31 satellites (in this very moment) distributed in six orbital planes, it is transposing at the global terrestrial scale old techniques of Euclidean geometry together with Newtonian physics. It essentially determines ranges between the observer and a number of satellites (normally six of them) by means of a time of flight determination of signals sent from the orbiting emitters. At the scale of the needed positioning it is immediately clear that special and general relativistic effects cannot be neglected, so they are introduced as corrections to the classical data. Relativity enters the process in order to take into account the behaviour of the orbiting atomic clocks: on one side their pace appears to be slowed down with respect to a similar clock at rest on the surface of the earth because of the orbital speed of the satellite; on the other side the frequency of the orbiting device is increased because of the gravitational blue-shift depending on the hight of the orbit. Finally the orbiting clocks must be synchronous with respect to one another and to the local clock of the user on earth, because of the need to measure times of flight. However the purely kinematical relativistic Sagnac effect produces de-synchronization of each clock with itself at each revolution, so that from earth one has to periodically re-align all clocks.

Besides these complications and the way they are managed, it is also true that GPS is not fit to guide spacecrafts navigating across the solar system. For such navigation other techniques are used, all requiring an almost continuous guidance from earth. The distance to the spacecraft can be determined with a good accuracy (in the order of millimeters) by means of laser or radio ranging from earth, but the transverse positioning is far worse and the accuracy rapidly decays with distance. Usually the spacecrafts for the exploration of the solar system are equipped with limited capacities of self-guidance; for instance they carry pictures of the sky that allow them to an autonomous control of their trim; similarly real or reconstructed images of the final destination enable the spacecraft to autonomously guess its distance and position with respect to the intended arrival. All this is however rather complicated and not easy to manage.

The above drawbacks, despite the enormous strength of political and commercial constraints, have begun to stimulate the search for a more up-to-date approach to positioning. To say the least, we speak today of space-time as a continuous four-dimensional Riemannian manifold with Lorentzian signature. We should try and start from that fact for the building of appropriate methods to be used in order to navigated across space-time; such an approach would include general relativity from scratch and not as a set of "corrections" to be made more or less by hands.

The consequent studies  have brought to a better definition of the concepts at the base of positioning, to the introduction of light coordinates and to the development of a number of proposals \cite{ref1}-\cite{ref8}. Here I shall present a relativistic positioning system that has been implemented to the level of algorithms and simulations \cite{ref9}-\cite{ref13}. It is based on the local measurement of the "length" (i.e. the proper time interval) of a stretch of the world-line of an observer between the arrivals of subsequent pulses from not less than four independent sources represented by known world-lines in space-time. An idea considered by many authors is the one of using pulsars \cite{ref14}-\cite{ref18},\cite{ref9}-\cite{ref11}, but other solutions can also be envisaged.

The timing, be it of laser or radio pulses, combined with relativistic positioning can also be of paramount importance for fundamental physics and I will shortly review some possibilities in the final part of the present document.

\section{Reference frames}
The very idea of finding a position in space-time implies the definition and assumption of a reference system with respect to which the position is defined. There can exist reference frames at various scales according to the peculiar applications one is interested in, however, in the end, some global frame needs be defined within which all other local and partial frames are located.
Of course what I am writing implies that a global reference frame \emph{can} indeed exist and uniquely be defined, which issue is not at all trivial when applied to the whole visible universe.

\begin{figure*}[]
\centering
\includegraphics[height = 60 mm, width=130 mm]{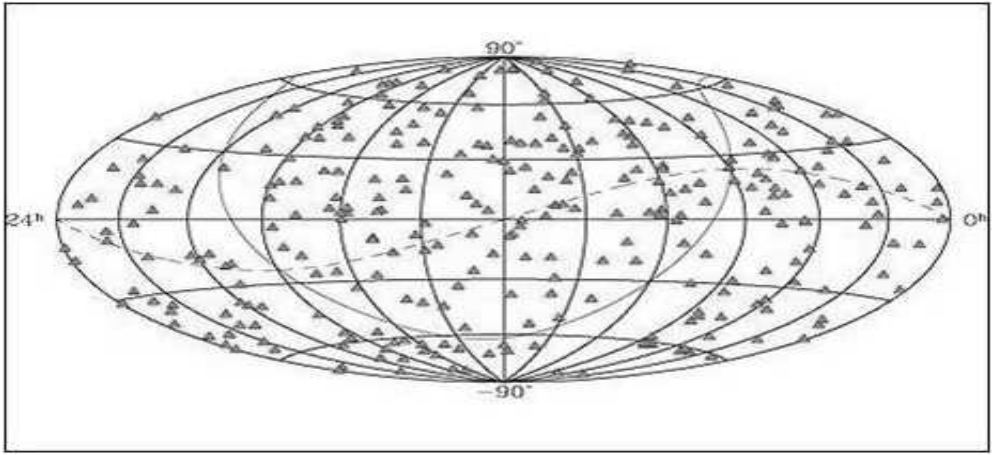}
\caption{The distribution in the sky of a few quasars taken from the Sloan Digital Survey}\label{sloan}
\end{figure*}

In practice the background reference frame that people commonly use is the one of the "fixed stars". Today by "fixed stars" quasars are meant. Quasars (quasi stellar objects) are, according to the most accepted interpretation, active galactic nuclei; the source of their energy is commonly ascribed to the presence of a massive black hole, but there are various hypotheses concerning the mass to energy conversion mechanism. What matters here, however, is that those bright object are very far away, from approximately 3 to approximately 13 billion light years. Their distance implies that, at the human time scales, the quasars appear as being fixed in the sky despite any proper motion they might be endowed with. From this fact arises the possibility of having fixed directions pointing along the axes of any Cartesian non-orthogonal reference frame. Thousands of quasars are known; fig. \ref{sloan} taken from the Sloan Digital Survey shows the distribution in the sky of a few of them. The reciprocal angular positions of the quasars in the sky are determined by the Very Long Baseline Interferometry (VLBI) and are known, at the moment, with an accuracy of the order of $10^{-9}$ rad.

Of course, in order to have a reference frame, three fixed non-co-planar directions are not enough. An origin also needs be chosen and this can be done arbitrarily, in principle. In practice the best we can do, at the moment, is to choose the origin of our frame located at the barycenter of the solar system. The barycenter of the solar system is indeed moving with respect to the quasars and its motion is not an inertial one because it is constrained by the gravitational field of the galaxy, but we assume that the acceleration due to the galaxy is negligible and consequently we pretend the motion to be inertial.

This essentially is the International Celestial Reference System (ICRS), being the quasars the International Celestial Reference Frame (ICRF).

We may then refer to the International Terrestrial Reference System (ITRS) which can be connected to the ICRS using the Earth Orientation Parameters given by the IERS (International Earth Rotation Service, now International Earth Rotation and Reference Systems Service).

\subsection{Space-time}

Everything I have written above is OK in three dimensions, however, if we wish to have a fundamental relativistic description, we must refer to space-time as a Riemannian continuum with Lorentzian signature. This means that we need to include time among the coordinates. When defining an origin for our reference system we need to introduce an origin of time as well; it can be arbitrary, of course, but what we really need is to associate a duration standard to our space origin. We consequently imagine to place an atomic clock in the barycenter of the solar system and to use its time as our coordinate time. This assumption is not trivial at all, since we know that, if we compare the readings of two identical clocks located along two different world-lines, we find they can differ from one another because of relative motion of the two clocks and them being placed in different gravitational potential wells.

\begin{figure*}[]
\centering
\includegraphics[height = 60 mm, width=130 mm]{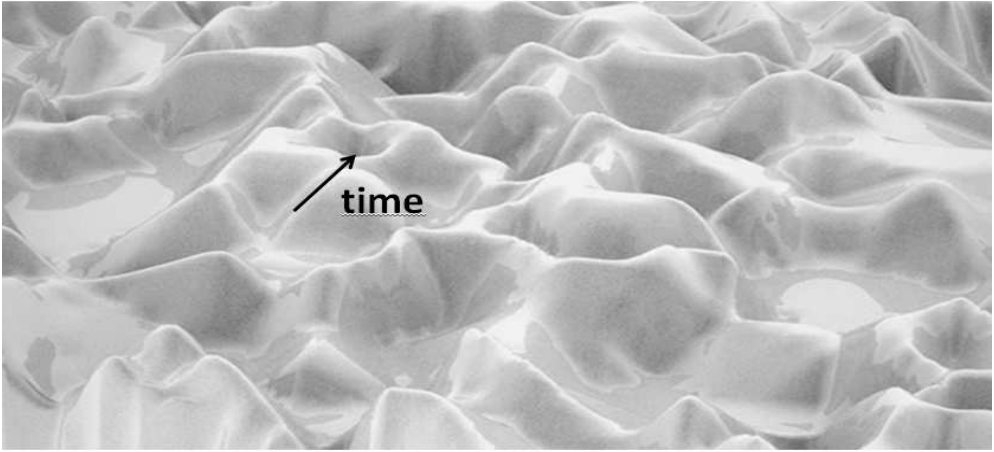}
\caption{Artistic three-dimensional view of a "crumpled" space-time}\label{crumple}
\end{figure*}

In fact, considering space-time, we see that it appears locally as a sort of crumpled manifold like in fig. \ref{crumple}.

The warps between the quasars and the local terrestrial observers can be no problem provided they stay stable during our progressive exploration of our world-line, over times of the order of typical human times. This assumption can be reasonable for the path along our past light-cone out of the solar system, but may be questionable in the final portion close to the observer, where the proper motion of the latter in the local gravitational potential well can introduce non-negligible time changes.

When setting up a global reference system in a general relativistic framework, our definitions actually rest on a number of implicit assumptions that are added to the explicit ones. For instance the ICRS implicitly assumes that space-time is asymptotically flat. Treating quasars as fixed (point-like) objects amounts to say that their world-lines are straight and parallel: this implies that "there" space-time is flat. Most likely, however, in the universe there is no asymptotic flatness; at most we may say that in between galaxies, far away from any matter bunch, space-time is \emph{almost} flat.

Putting everything together, we should rather say that the ICRS (or any other analogous reference system) is defined as being "drawn" on a flat Minkowski space-time which coincides with the tangent space-time in the origin of our reference frame. In order to use the proper time of the atomic clock located in the origin as the global coordinate time of our reference implies that we treat it as being in a globally flat environment.

We have no such problem at the moment, but the use of the local tangent space-time would make it not trivial at all to uniquely and understandably transfer position information to another observer a few million light years away.

\subsection{Coordinates and geodesics}

\begin{figure*}[]
\centering
\includegraphics[height = 78 mm, width=130 mm]{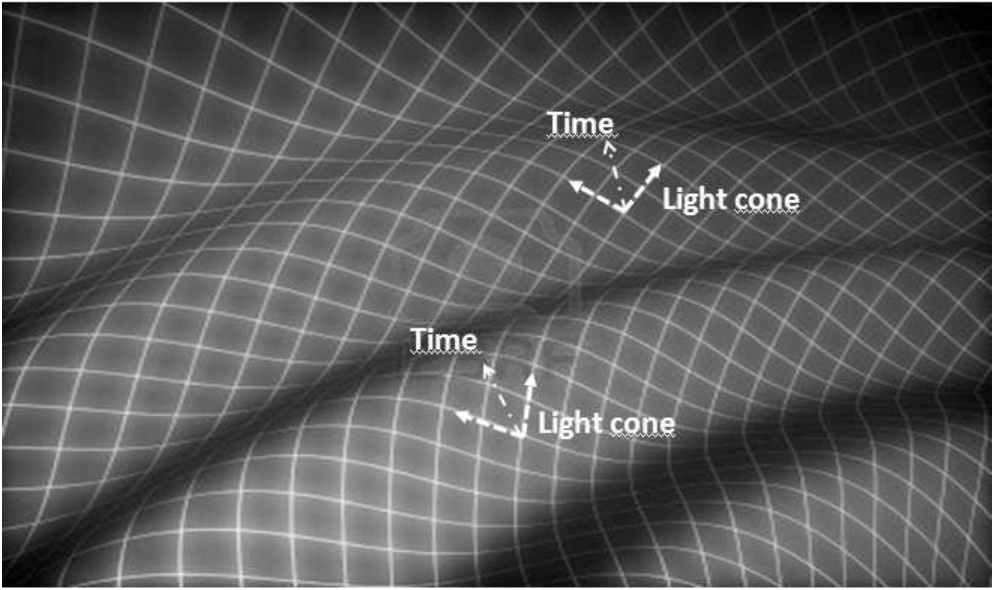}
\caption{Two sets of null geodesics covering a bidimensional curved manifold}\label{nullgeo}
\end{figure*}

Looking at the problem of defining efficient reference systems for a Riemannian manifold we need also to decide how to uniquely and smoothly attribute to each event in the manifold a quadruple of numbers i.e. an appropriate coordinate set. This may be done in principle drawing four independent families of curves. The curves of each family do not intersect each other and densely cover the whole manifold. Labeling each curve by a progressive real number, any intersection of four curves from different families identifies an event on the manifold and endows it with four coordinates. What I have described by this process is known as a Gaussian coordinates system.
As for the curves to be used they can for instance be geodesics of the manifold. Not considering singularities and defects, geodesics do indeed cover the whole manifold.
In the case of space-time and in view of the importance of the null cones a good choice for building a Gaussian coordinate system is to use independent families of null geodesics. Fig. \ref{nullgeo} visualizes our choice in a simplified bi-dimensional space-time.

In a flat manifold a whole family of null geodesics may be identified by means of a tangent vector to anyone of them. It would be a null four-vector like the one written in eq. (\ref{eq:null}).

\begin{equation}
\label{eq:null}
\chi=cT(1,\cos\alpha,\cos\beta,\cos\gamma)
\end{equation}

It is indeed $\chi^2=0$. The space components of (\ref{eq:null}) are the direction cosines of a three-vector defined with respect to some global reference system of our choice. The $cT$ factor is optional and does not change the null character of $\chi$, but contains an additional information on the period of the signals traveling along the null geodesics of the family identified by $\chi$, provided they are indeed periodic.

Suppose now that we have four independent null four-vectors $\chi_{1,2,3,4}$ and use them as a basis for vectors in the manifold; the position of any event in the same manifold is identified by a "radial" four-vector, expressed as a linear combination of the basis vectors:

\begin{equation}
\label{eq:vector}
r=(\frac{\tau}{T})_a\chi^a
\end{equation}

The index $a$ runs from 1 to 4; the periods of the signals are assumed to be different from one another in the quadruple; $\tau$'s are named "light coordinates" of the event and the ratios $(\frac{\tau}{T})_a=x_a$ are pure numbers due to the choice for the scale factor in eq. (\ref{eq:null}).

The number of actual degrees of freedom in our representation deserves a comment. All $\chi$'s of the basis are on the light cone of the origin of the reference frame (actually on any light cone if the manifold is globally flat), which means that, as far as we stay on the light cone, only three of the null wave vectors can be mutually independent. Three $\chi$'s are enough to localize events on the light cone; we need four of them for time-like or space-like events, i.e. for events out of the light cone.

Rather than using the families of null geodesics, we may adopt a dual vision. Each $\chi$ is associated with a null four-dimensional hyperplane obtained by Hodge conjugation. The corresponding four-form $\varpi$ is $\varpi=\ast\chi$, or explicitly:

\begin{equation}
\label{eq:omega}
\varpi_{abc}=\epsilon_{abcd}\chi^d
\end{equation}

$\epsilon_{abcd}$ is the Levi-Civita fully antisymmetric tensor. Now we have four independent families of hyperplanes covering the whole space-time and intersecting each other. The hyperplanes of a family are null and orthogonal to the corresponding $\chi$.

All this is globally true if the manifold is flat; if it is curved it holds locally.

\section{Positioning}

On the bases laid down in the previous section, we may outline a fully relativistic positioning method.
Suppose you have (not less than) four independent sources of electromagnetic signals located at infinity; suppose then that they emit pulses at the rate of $1/T$ per second. The $T$ parameter of formula (\ref{eq:null}) is now interpreted as the repetition time of the pulses rather than the period of a monochromatic continuous wave. Once this has been specified we may apply the procedure outlined in the previous section. The $\chi$'s are associated to the four (or more) sources; we may identify as duals to the $\chi$'s four discrete sets of hyperplanes $\varpi$ covering space-time with an egg crate whose spacings along the directions of the basis vectors are given by the $T$'s, when measured along the time axis of the background global reference frame.

\begin{figure*}[]
\centering
\includegraphics[height = 70 mm, width=120 mm]{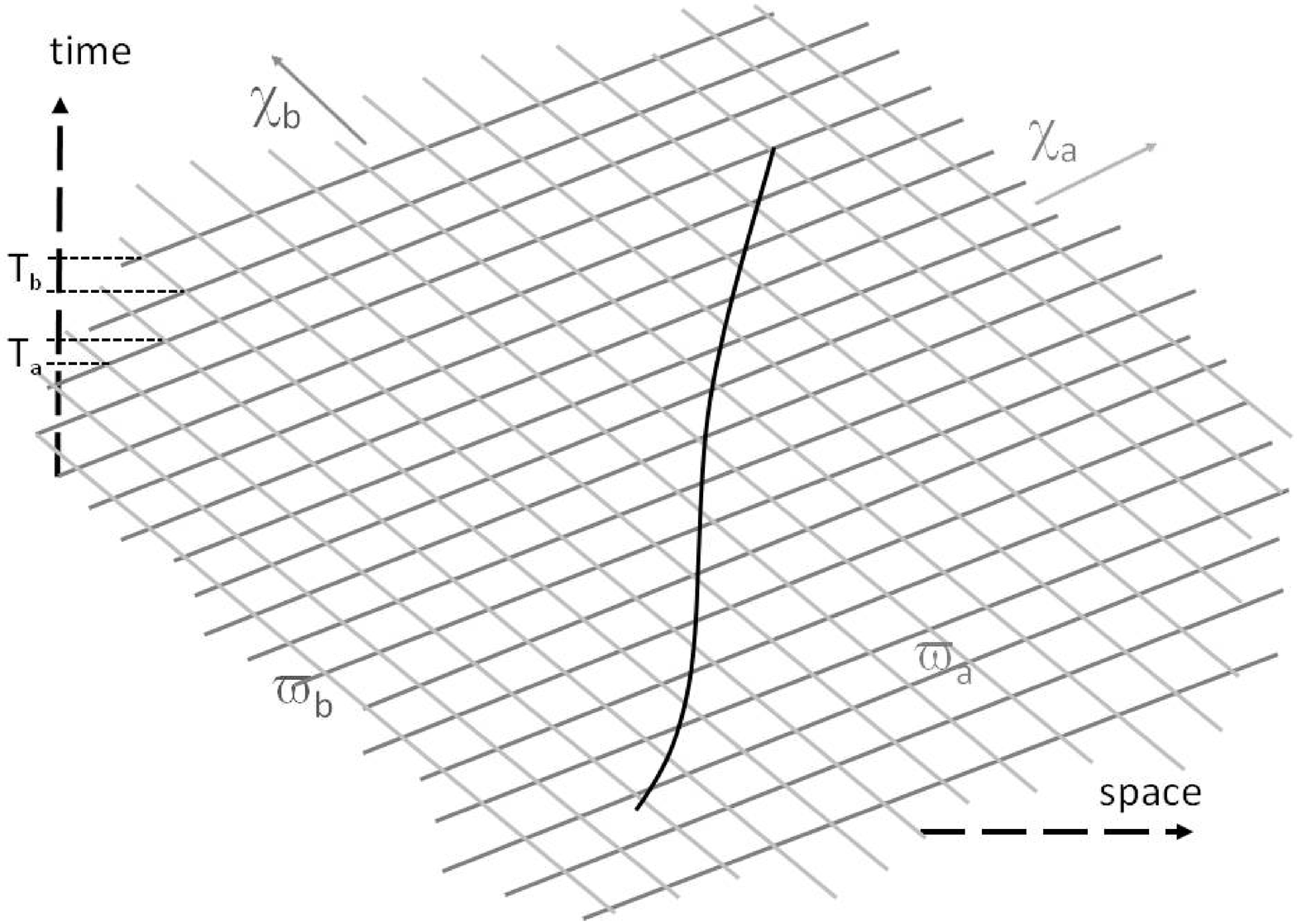}
\caption{A bidimensional flat space-time covered by a grid made of null hypersurfaces (actually lines) conjugated to the null vectors $\chi_{a,b}$. The wavy line is the world-line of an observer.}\label{grid}
\end{figure*}

The situation is schematically shown in fig. \ref{grid}.

\begin{figure*}[]
\centering
\includegraphics[height = 80 mm, width=110 mm]{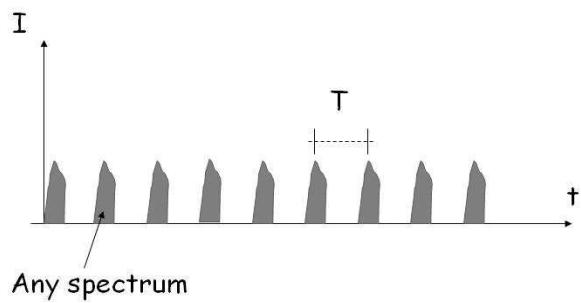}
\caption{Typical emission sequence of the pulses from a source. Vertically intensities are drawn; the profile of the pulse is not important; times are proper times of the emitter.}\label{pulse}
\end{figure*}

The world-line of an observer necessarily crosses the walls of successive boxes of the egg crate. If we are able to label each cell of the crate, we are also able to reconstruct the position of the observer in the manifold. The use of pulses implies that, realistically, the walls of the cells are "thick". In practice the hypersurfaces on the graph correspond to "sandwich waves" carrying the pulse. A typical emission diagram of one of the sources will more or less be like the one sketched in fig. \ref{pulse}.

The shape of the pulse is not important as well as it is not the spectral content of it. What matters is its reproducibility and the stability of the repetition time. Considering natural pulses, as the ones coming from pulsars, we find repetition times ranging from several seconds down to a few milliseconds and lasting a fraction of the period. As an example of artificial pulses the highest performance is obtained with lasers: GHz frequencies are possible with pulses as short as $\sim10^{-15}$ s.

\begin{figure*}[]
\centering
\includegraphics[height = 80 mm, width=110 mm]{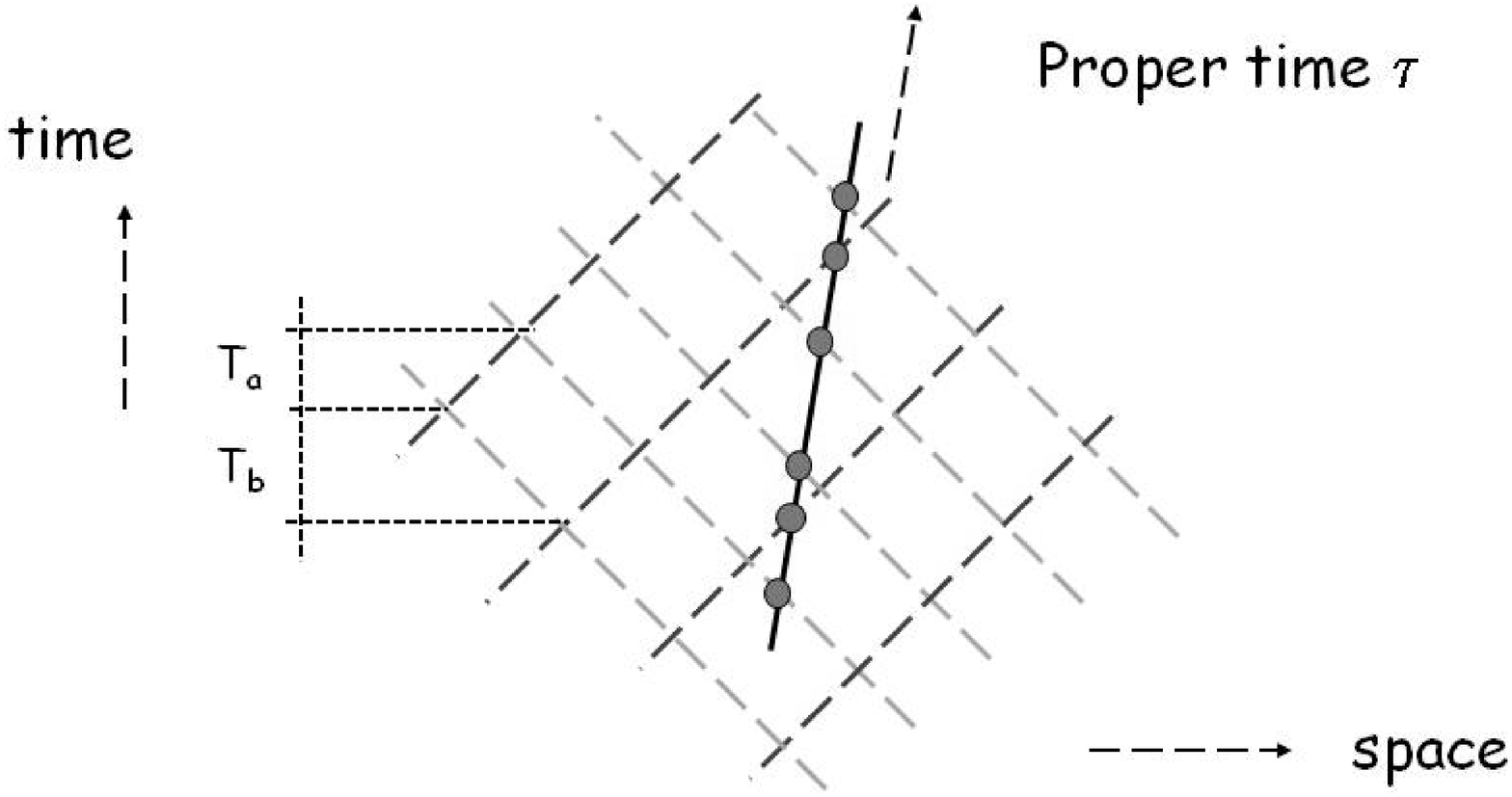}
\caption{A straight portion of a world-line is shown in a local flat patch of space-time. The lines of the grid correspond to different pulses labeled by their ordinal integer. The intersections of the world-line with the walls are localized by quadruples (actually pairs in the figure) of real numbers one of which is always an integer.}\label{flat}
\end{figure*}

Once pulses are used, we may label them in order, by integer numbers, as it can schematically be seen in fig. \ref{flat}.

The integers can be though of as rough coordinates identifying the cells of the grid. At this level the approximation would be rather poor, being of the order of the size of each cell. If the periods are of the order of milliseconds this corresponds to hundreds of kilometers. Looking at fig. \ref{flat} we may however notice that the intersections of a given world-line with the walls of the cells are labeled by a quadruple of numbers, at least one of which is an integer: these numbers are the coordinates of the intersection points. We may write the typical coordinate of a position in the crate as $\xi_a=n_a+x_a$; the $n$'s are the integers, whilst the $x$'s are the fractional parts. If we have a means to determine the $x$'s the localization of an intersection event can be done with an accuracy much better than the hundreds of km I mentioned above. Considering that the intersections coincide with the arrivals of pulses from different sources, the determination of the fractional part of the coordinates is indeed a trivial task, provided the traveler carries a clock, the space-time is flat and the world-line is straight. Once one measures the proper intervals between the arrivals of successive pulses a simple linear algorithm based on elementary four-dimensional flat geometry produces the $x$'s \cite{ref11}. The corresponding light coordinates are $\tau_a=[(n+x)T]_a$. The accuracy of the result depends on the precision of the clock which is being used in order to measure the proper intervals between pulses and on the stability of the period of the pulses, which in turn tells us what the effective "thickness" of the walls of the cells of our space-time crate is. Just to fix some order of magnitude, let me remark that nowadays to have a portable clock with a $10^{-10}$s accuracy is quite easy (much better can be achieved in the lab); on the other side, considering pulsars, we have some, whose period is known and stable down to $10^{-15}$s. With these figures the final positioning can be within a few centimeters.

Of course the traveler's motion will not in general be an inertial one and space-time will not be flat, however a short enough stretch of the world-line can always be confused with the tangent straight line to it and a small enough patch of space-time can always be confused with a portion of the local tangent space. In practice we work on the local tangent space and on a linearized portion of the world-line. The acceptability of these assumptions depends on the accuracy required for the positioning and on the constraints posed by the linear algorithm in use. The reconstruction of a piece of the world-line requires the knowledge of at least eight successive arrival times of pulses from the minimal set of independent sources (four) \cite{ref11}. So, if $\delta \tau$ is the maximum proper time inaccuracy that we decide to be tolerable, the final relative accuracy of the positioning will be:

\begin{equation}
\label{eq:delta}
|\frac{\delta x}{x}|\leq 4(\frac{1}{\tau_{i,i+4n}}+\frac{\tau_{i,i+1}}{\tau_{i,i+4n}^2})\delta\tau
\end{equation}

The index $i$ in eq. (\ref{eq:delta}) labels the order of the arrival events; $\tau_{i,i+4n}$ is the proper time interval between the $i$th and the $(i+4n)$th arrival, being $n\geq1$ an integer; $n$ should assume the highest value compatible with the straightness hypothesis for the world-line. Of course the number of pitches that can safely be considered depends on the periods $T_a$ of the emitting sources: the shorter are the periods, the bigger is the number of paces that can be used within the linearity assumption.

A pictorial view of what we are doing is as follows. Imagine to embed the real four-dimensional manifold, together with its tangent space at the start event, in a five-dimensional flat manifold; then consider the real world-line of the traveler and project it onto the tangent space. The world-line on the tangent space is what we are piecewise reconstructing by our linear algorithm: in practice we are building a flat chart containing the projection of our space-time trajectory. The time dependence of the a-dimensional coordinates of the projected world-line may of course be written in the form of a power series, as:

\begin{equation}
\label{eq:serie}
x_a=u_a\frac{\tau}{T_a}+\frac{1}{2}\alpha_a\frac{\tau^2}{T_a^2}+...
\end{equation}

The coefficients $u_a$ and $\alpha_a$ are proportional to the four-velocity and four-acceleration of the traveler. The individual segments used for the reconstruction are short enough so that the second and further terms of (\ref{eq:serie}) are negligible with respect to the linear one. Going on, after a number of paces, the possible presence of an extrinsic curvature of the projected world-line shows up; we know that locally it is impossible to distinguish a gravitational field from a non-gravitational acceleration so we need additional information for that purpose. In the case of a gravitational field evidenced by the reconstruction process I am describing, we get from the data the gradient of the Newtonian gravitational potential.

In order not to cumulate the distortion introduced by the projection from the real curved manifold to the tangent space at a given event, we need periodically to restart from a further event on the world-line, i.e. to pass to the tangent space at a different event. If the visible curvature of the line on the tangent space as well as the tilt of the successive tangent spaces continues for long in the same sense, the linearization process, as in all similar cases, tends to produce a growing systematic discrepancy with respect to the real world-line, so that periodically one has to have recourse to some independent position fixing means in order to reset the procedure.

\subsection{Pulsars}

I have already mentioned pulsars as possible natural sources of pulses. This kind of neutron stars are indeed good pulse emitters because of their extreme stability and long duration. As we know, their emission is in the form of a continuous beam. The apparent periodicity is due to the fact that the emission axis (the magnetic axis) does not coincide with the spin axis of the object so that it steadily rotates, together with the whole star, about the direction of the angular momentum. The pulses arise from the periodic illumination of the earth by the rotating beam. The stability is guaranteed by the angular momentum conservation.

The advantages of pulsars are numerous. Their period is extremely stable and is sometimes known with the accuracy of $10^{-15}$ s; it tends to decay slowly (the relevant times are at least months), but with a very well known trend, determined by the emission of gravitational radiation. Typically the fractional decay rate of the period is in the order of one part in $10^{12}$ per year. The number of such sources is rather high, so that redundancy in the choice of the sources is not a problem: at present approximately 2000 pulsars are known and their number continues to increase year after year. Being these stars at distances of thousands of light years from the earth, they can be treated as being practically fixed in the sky; in any case their slow apparent motion in the sky is known, so corrections for the position are easily introduced. Just to recall some numbers, the rate of change of a typical angular coordinate $\alpha$  in the position in the sky is

\begin{equation}
\label{eq:alpha}
|\frac{\delta\alpha}{t}|\approx 10^{-6} (\frac{100 pc}{distance})\frac{rad}{year}
\end{equation}

Unfortunately pulsars have also major drawbacks. One is that their distribution in the sky is uneven, since they are mostly concentrated in the galactic plane, which fact brings about the so called "geometric dilution" of the accuracy of the final positioning: sources located on the same side of the observer produce an amplification of the inaccuracy originating in the intrinsic uncertainties. Furthermore individual pulses differ in shape from one another so that some integration time is needed in order to reconstruct a fiducial series of pulses; this fact, also considering the length of the repetition time, can conflict with the linearization of the world-line of the traveler. It should also be mentioned that most pulsars are subject to sudden jumps in the frequency (glitches), causes by matter falling onto the star; these unpredictable changes can be made unoffensive by means of redundancy, i.e. making use of more than four sources at a time.

However the most relevant inconvenience with pulsars is their extreme faintness. In the radio domain their signals can be even 50 dB below the noise at the corresponding frequencies; to overcome this problem big antennas are required (not less than 100 $m^2$) and convenient integration times accompanied with "folding" techniques must be employed. In principle at least four different sources must be looked at simultaneously and this is not an easy task, especially with huge antennas.
The weakness problem has led to consider X-ray- rather than radio-pulsars for positioning. A few hundreds X-ray emitting pulsars are indeed known; their signals are weak too, and can be received only outside the atmosphere, but the background noise is far smaller than the one typical in the radio domain; as for the hardware, X-ray antennas can be much smaller than the typical radio-antennas. Since many galactic X sources emit also at radio frequencies, one can envisage the opportunity to combine both X-ray and radio pulses from one single source for the positioning process.

\subsection{Artificial and blended solutions}

In principle what can be done using pulsars can as well be done by means of artificial emitters of electromagnetic pulses. Artificial emitters can have far higher intensities than pulsars; the repetition time can easily be in the range of ns or less, thus making the linearization process more reliable. The stability over time of the source is not as good as for pulsars, but this can represent no inconvenience as far as the number of sources is redundant and they are kept under control. A problem is in the sources clearly not being at infinite distance, which implies a more complicated geometry and of course the need for a good knowledge of the world-line of the emitter in the background reference frame.

One could think of building a Solar System reference frame made of pulse emitters laid down on the surface of various celestial bodies whose orbits are well known and reproducible: the earth of course, the moon, Mars, maybe some of the asteroids; even some space station following a well defined, highly stable orbit around the sun or a planet.

A blended solution for self-guided navigation in the solar system could combine some artificial emitters, as quoted above, together with a limited number of pulsars (the most intensely emitting ones).

The fully relativistic method I have described is of course specially fit for space navigation, but it can also be useful in rather limited areas. Think for instance to the accurate mapping of a depopulated region, where traditional topography may be rather expensive. If one puts a limited number of antennas (not less than four in any case) emitting pulses, located in precisely defined positions at the boundary of the region to map, their signals may be used by a moving vehicle to draw a chart of the area within centimeter accuracy. Of course the same can be done using differential GPS, but with the specific features of GPS, mainly the fact that it is under military control and the best performance of the system is not reserved to ordinary civilian applications.

One could also think that the new relativistic method will be implemented in the next generations of global positioning systems, even though the approach tends to be rather conservative there, the reason being that the huge amount of money already spent for developing and deploying the traditional GPS makes present applications based on it cheap, whereas any new solution would initially be more expensive. Probably a gradual transition will happen, triggered by new applications, especially outside the terrestrial environment, and, last but not least, political reasons.

\section{Positioning and fundamental physics in space}

The method I have being describing for positioning purposes is based on electromagnetic signals and their accurate timing. The same kind of technology can be used for various experiments aimed to the detection of fundamental properties of space-time. It is worth mentioning a few possibilities.

\subsection{Intercommunicating swarms of satellites}

Consider a swarm of identical satellites (as the ones of the future Galileo system), equipped with pulse emitters and receivers and able to accurately measure the arrival times and to recognize the origin of each pulse (this could be achieved tuning the emitters on different individual frequencies). The information gathered by the whole constellation would allow for space-time geodesy, based on  multiple triangulations performed on null triangles. It would be a means to reconstruct the average curvature of the patch of the manifold where the world-lines of the satellites lie. The Galileo satellites will indeed be able to intercommunicate and also the present GPS satellites may communicate with each other even though this possibility (introduced for military reasons) is not actually used at the moment.

\subsection{Ring-lasers}

Electromagnetic waves can be used as probes for the structure of space-time and in particular the gravito-magnetic part of the gravitational interaction, exploiting the anisotropic propagation of light induced by the chiral symmetry associated with a rotating mass. This possibility is the basis of the proposal to use ring lasers for the measurement of the Lense-Thirring (frame dragging) effect of the earth \cite{ref19}. If a light beam is obliged, by conveniently located mirrors, to follow a closed path in space, the total time of flight for a loop is different according to the fact that light is moving in the same sense as the rotation of the central mass or in the opposite sense. In fact the difference in the proper (i.e. of the laboratory) times of flight for one turn in co- and counter-rotating sense, is obtained as \cite{ref19}:

\begin{equation}
\label{eq:tof}
\delta\tau = -2\sqrt{g_{00}}\oint{\frac{g_{0\phi}}{g_{00}}d\phi}
\end{equation}

Polar coordinates centered on the earth are assumed and the $g_{\mu\nu}$'s are elements of the metric tensor of an axially symmetric stationary space-time. A ring laser converts the time of flight difference in a beat note obtained from the two counter-rotating beams in steady state; the beat note arises from the different equilibrium frequencies of the two beams. The device in practice can measure effective angular velocities, which contain the kinematical effect (classical Sagnac effect), the geodetic (or de Sitter) effect (coupling of the gravito-electric field of the earth with the kinematical rotation of the apparatus) and the gravito-magnetic contribution (Lense-Thirring frame dragging). The latter two terms are 9 orders of magnitude smaller than the classical Sagnac effect when measured on the surface of the planet, so that a very high sensitivity is needed, but contemporary laser technologies are approaching the required accuracy level. Both the de Sitter and Lense-Thirring effects have already been measured in space by a different technique based on the behaviour of mechanical gyroscopes. The most difficult to reveal is the Lense-Thirring drag and it has been measured with an accuracy of $19\%$  by the GP-B experiment \cite{ref20} and $10\%$ by the laser ranging of the LAGEOS satellites \cite{ref21}; laser ranging of the orbit is also being used by the LARES experiment launched on February 13th 2012, with the purpose of reaching the $1\%$ accuracy \cite{ref22}. The newly proposed ring laser GINGER experiment \cite{ref19} is aimed at reaching a $1\%$ accuracy for the physical terms in a terrestrial laboratory.

Here I would like to mention the possibility of bringing a ring laser experiment in space. One could for instance think of a three-dimensional array of four mirrors, rigidly attached to one another in the shape of a tetrahedron; the whole thing could be in free fall (stable orbit) around the earth. Each face of the tetrahedron would coincide with a triangular ring laser; the signal extracted from each face would give information on the projection of the total rotation vector on the normals to the faces. In the case of a circular equatorial orbit the frequency of the beat note extracted from one of the faces of area $S$ and perimeter length $P$ would be:

\begin{eqnarray}
\label{eq:tri}
f_{beat} &=& 4\frac{S}{\lambda PR}\sqrt{G\frac{M}{R}}\nonumber\\
&\times&[(1-\frac{3}{4}\sqrt{G\frac{M}{c^2R}} + 4G\frac{M}{c^2R})\hat{n_a}\cdot\hat{n_S}\nonumber \\
&-&(\frac{1}{2}G\frac{M}{c^2R}+\frac{GJ}{c^4R})\hat{n_\theta}\cdot\hat{n_S}]
\end{eqnarray}

$M$ is the mass of the earth; $J$ is its angular momentum; $G$ is Newton's constant; $R$ is the radius of the orbit; $\lambda$ is the wavelength of the light of the laser; $\hat{n_S}$, $\hat{n_a}$, $\hat{n_\theta}$ are unit vectors, respectively, perpendicular to the plane of the ring, aligned with the axis of the earth, aligned with the local meridian in the sense of increasing co-latitude (here, in practice, perpendicular to the equatorial plane).

The term depending on the angular momentum of the earth in eq. (\ref{eq:tri}) is the smallest and is eight orders of magnitude below the biggest; an extremely good accuracy is always required, but in free fall one has a far smaller environmental noise than on earth.

\subsection{Linear cavities}

\begin{figure*}[]
\centering
\includegraphics[height = 80 mm, width=110 mm]{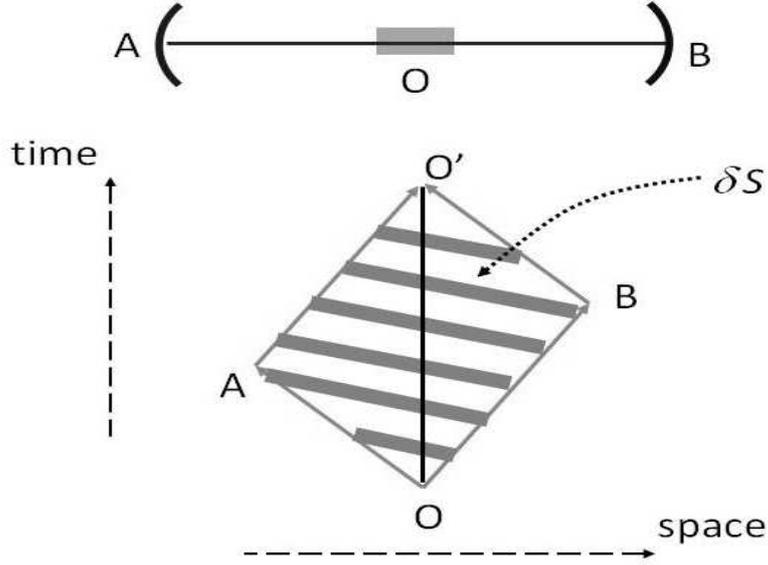}
\caption{World-sheet representation of light being reflected back and forth in a linear cavity. A and B are mirrors; O-O' is the world-line of  an active region in the middle of the two mirrors; the gray stripes identify a space-time area $\delta S$.}\label{cavity}
\end{figure*}

Another interesting possibility is represented by simple linear resonating cavities as the ones in Fabry-P\'{e}rot interferometers. In fact, when describing a simple bounce back from a mirror to the other in four-dimensional space-time, one has a bidimensional graph, like the one shown on fig. \ref{cavity} where an active region is assumed in the middle of two mirrors.

As it can be seen, the light beams moving back and forth in the cavity delimit a closed contour at each cycle. This fact implies that the effects of the curvature and the chiral symmetry of space-time may expressed in terms of the Riemann tensor and the contoured area. Considering the electromagnetic tensor $F_{\mu\nu}$, its change after one cycle is given by:

\begin{equation}
\label{eq:cav}
\delta F^{\mu\nu} = (R^{\mu}_{\epsilon 0i}F^{\epsilon\nu}+R^{\nu}_{\epsilon 0i}F^{\mu\epsilon})\delta S^{0i}
\end{equation}

Latin indices are used for space-coordinates; $\delta S^{\mu\nu}$ is the antisymmetric area 2-form.

For practical purposes, only the least useful approximation of the Riemann tensor needs be retained down to the order of the angular momentum of the earth. An example of the approximated version of one of the equations (\ref{eq:cav}) is for instance:

\begin{equation}
\label{eq:exam}
\delta F^{\theta\phi} \cong (\frac{GM^{3/2}}{c^3R^{7/2}}-3\frac{GJ}{c^3R^4})\frac{\cos{\theta}}{\sin^2{\theta}}\frac{l^2}{R}F^{\theta r}
\end{equation}

$\delta F^{\theta\phi}$ is the change in the radial component of the magnetic field expressed as a function of the East-West component; $l$ is the length of the cavity and $R$ is its radial position with respect to the center of the earth.

The result depends on the orientation of the cavity and builds up with the successive reflections. One could think of combining this effect along an array of mutually perpendicular freely falling cavities.

\begin{figure*}[]
\centering
\includegraphics[height = 80 mm, width=110 mm]{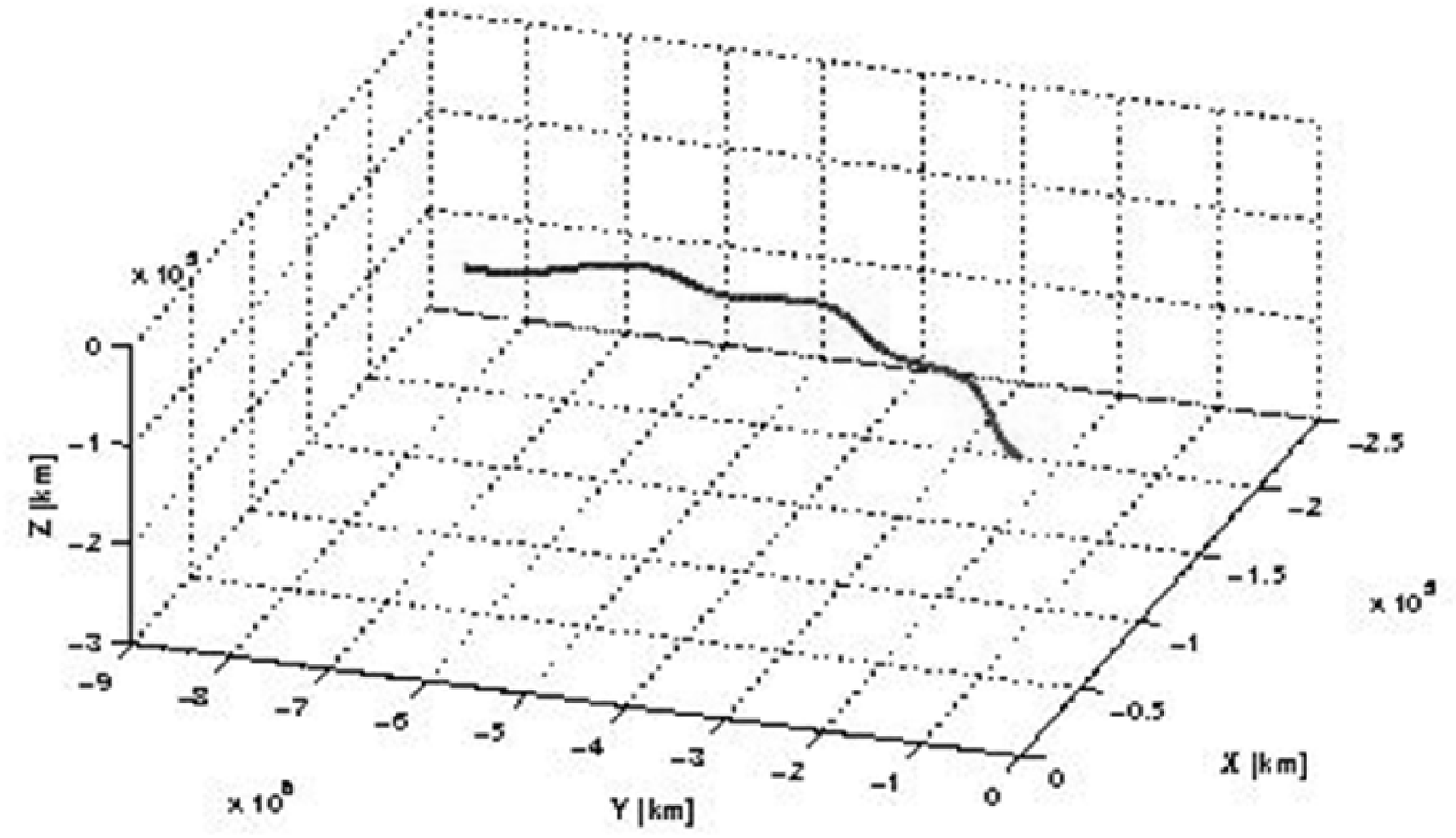}
\caption{The picture shows the motion of the Parkes observatory carried by the earth in its rotation and revolution motion, during three days. The reconstruction has been made applying the relativistic positioning method to the simulated arrival times of the signals from four real pulsars. The reconstructed trajectory is superposed to the real one.}\label{eppur}
\end{figure*}

\subsection{An orbital ring cavity}

The satellites of the GPS constellation are distributed on 6 different orbital planes so that 5 of them are on the same orbit; the Galileo system, when fully deployed will have 30 satellites on 3 orbital planes so that 10 satellites will share the same orbit. The presence of at least three satellites on the same orbit opens an interesting possibility if they are enabled to communicate with each other. Suppose each satellite is sending laser pulses to the others who are forwarding them along the orbit. In practice we would have a sort of ring laser at orbital scale. If at least one of the satellites is equipped with interferometric devices or can accurately measure the arrival time difference between pulses having completed a clockwise turn with respect to the ones revolving in the counter-clockwise sense, the whole system behaves a gigantic ring laser (or Sagnac interferometer) with a sensitivity measured by the huge scale factor given by the ration between the contoured area and the length of the perimeter of the polygon followed by the light pulses.

\section{Conclusion}

Summing up, I have shown how the approaches to positioning and navigation could be implemented in order to become fully relativistic. The idea is in the use of space-time as such as a reference, and in the exploitation of four-dimensional geometry. In practice a generalization is possible of the ordinary three-dimensional topographic techniques, upgrading them to four dimensions and the Lorentzian signature. The method, with the related algorithms, is per se simple and relies on plain proper time measurements made by the traveller needing to localize himself in a given background reference frame. Reducing everything to the essentials, we see that the pattern of the proper arrival times of regular pulses from not less than four independent sources is uniquely related to the position of the receiver in space-time and to its starting event.
The relativistic positioning method has positively been tested with simulators. Fig. \ref{eppur} shows for instance the reconstruction of three days of the absolute motion of the antenna of the radio-telescope at the Parkes observatory,obtained by the simulated timing of four real pulsars \cite{ref10}.

This approach, either based on signals from pulsars (X ray or radio waves emitters) or on artificial sources laid down on the surface of the earth and other bodies of the solar system, will probably raise growing interest little by little as the need for positioning systems freed from the control of any specific power will increase. The same holds with the expansion of navigation within the solar system.
I have also given a few examples of the importance of timing measurements together with the use of laser beams or pulses for fundamental physics. It turns out that light is indeed a perfect relativistic probe for testing the structure of space-time.

%\bibliographystyle{plain}
%\bibliography{tartaglia}

\end{document}